\documentclass[12pt, nofootinbib]{revtex4-2}
\usepackage{amsmath,amssymb,bbold,bm}
\usepackage{caption}
\usepackage{hyperref}
\usepackage{float}
\usepackage[makeroom]{cancel} 
\usepackage{braket}
\usepackage{booktabs}
\usepackage{graphicx}
\usepackage{subcaption}
\usepackage{tablefootnote}

\usepackage{xcolor}
\usepackage{ulem}

\newcommand{\be}{\begin{equation}}
\newcommand{\ee}{\end{equation}}
\newcommand{\ba}{\begin{align}}
\newcommand{\ea}{\end{align}}
\newcommand{\bea}{\begin{eqnarray}}
\newcommand{\eea}{\end{eqnarray}}

\begin{document}
\title{To Infinity and Beyond: Maximizing Entanglement and Emergent Symmetries in $N$ Higgs Doublet Models}
\author{Spencer Chang, Gabriel Jacobo}
\address{
Department of Physics and Institute for Fundamental Science\\ 
University of Oregon, Eugene, Oregon 97403}
\begin{abstract}
In this paper, we explore the recent observation that maximizing flavor entanglement in Higgs boson scattering in a two Higgs doublet model leads to parameter space where the potential, if bounded from below, has an enhanced mirror symmetry $U(2)^2$.  Looking at $N$ Higgs doublet models, we find a generalization of the mirror models, with $U(2)^N$ symmetry.  Focusing on three Higgs doublet models, we determine that there is also a bounded potential solution that maximally entangles but has only the minimal $U(2)$ symmetry.  This counterexample demonstrates that additional constraints must be imposed to maintain the connection between maximal entanglement and extended symmetries.  As a potential constraint, for the three Higgs doublet model, we show that requiring maximal entanglement still occurs after gauging the symmetries eliminates the counterexample, thus restoring the connection.
\end{abstract}

\maketitle
\section{Introduction}
Symmetry plays a central role in particle physics, enabling deep insights into the allowed particles, their masses, and interactions.  Given this importance, it's interesting that there have been recent examples where extremization of the entanglement produced in scattering has been shown to select parameter space with enhanced symmetries.  This gives the hope that a quantum information perspective can give additional tools to diagnose/understand symmetries. 

The potential connections between symmetry and {\bf minimizing} entanglement has been explored in  low energy hadron scattering \cite{Beane:2018oxh, Low:2021ufv, Liu:2022grf, Hu:2024hex, Sone:2026jmo}, the two Higgs doublet model \cite{Carena:2023vjc,Chang:2024wrx, Kowalska:2024kbs, Kowalska:2025qmf}, and in general \cite{McGinnis:2025brt, Kanemura:2026won}.  However, there is typically a tension when minimizing entanglement, due to the fact that a free theory is minimally entangling.  In fact, it is challenging for many scattering channels to minimize entanglement, so imposing this constraint to all channels will generically lead to a highly symmetric, yet trivial, free theory \cite{Chang:2024wrx}.  

More recently, it has been studied if {\bf maximizing} entanglement is related to symmetric parameter points \cite{Carena:2025wyh, Li:2026kha}.  In the two Higgs doublet model, it has been shown \cite{Carena:2025wyh} that maximizing entanglement leads to a mirror $U(2)^2$ symmetry, which can even be extended to gauging the symmetries and adding Yukawa interactions with fermions.  Notably, all two to two channels can maximize entanglement with no danger of reducing to a trivial theory.\footnote{There has also been a research direction, trying to use entanglement extremization to determine parameter points close to experimental values, e.g.~\cite{Cervera-Lierta:2017tdt,Thaler:2024anb,Liu:2025iwh}.}  

In this paper, we further investigate this connection between maximizing entanglement and symmetry, by trying to extend \cite{Carena:2025wyh} to a general $N$ Higgs doublet model.  Interestingly, we find coupling choices that generalize the mirror models, with a $U(2)^N$ symmetry, that maximize entanglement.  We then look in detail at the three Higgs doublet model to see if this is the only viable solution.  As in the two Higgs doublet model \cite{Carena:2025wyh}, we find many solutions which maximize entanglement, but most are eliminated after requiring the potentials to be bounded from below.  However, we also find an entanglement maximizing solution that is stable, but only has the minimal $U(2)$ symmetry.  Thus, for this connection between maximal entanglement and symmetry to be reliable, additional constraints, aside from stability and maximizing entanglement, need to be imposed to restore the connection.  As a possibility of such a constraint, we demonstrate that requiring the maximal entanglement to still hold with the symmetries being gauged eliminates the counterexample, leading to a proposed refinement of the requirements for the connection between maximizing entanglement and extended symmetries.  

The layout of the rest of this paper is as follows.  
In Sec.~\ref{Sect:Procedure}, we discuss how to quantify entanglement for the two to two scattering channels of our models.  In Sec.~\ref{Sect:2HDM}, we reanalyze the two Higgs doublet models, both as a review and to confirm our analysis can reproduce the results in \cite{Carena:2025wyh}.  In Sec.~\ref{Sect:NHDM}, we describe our proposed generalization of the mirror symmetry to the $N$ Higgs doublet model and confirm that its scattering maximizes entanglement and the potential is bounded from below. In Sec.~\ref{Sect:3HDM}, we study the 3 Higgs doublet model in detail, finding all of the entanglement maximizing solutions whose potential is stable.  As described above, we find both the generalized mirror model, but also a coupling structure which has a stable potential but no enhanced symmetries. We demonstrate that requiring entanglement maximization still occurs after gauging the symmetry eliminates the counterexample, leading to a potential refinement of the general correspondence between entanglement maximization and symmetries.  Finally, in Sec.~\ref{Sect:conclusions}, we conclude and give some future outlook.

\section{Maximal Entanglement in $2\to 2$ scattering}\label{Sect:Procedure}
In this section, we present the framework used to analyze the entanglement generated by the scattering amplitude.
For an initial two body state $\ket{\text{initial}}=\ket{ij}$ comprised of particles from sectors $A$ and $B$, a two body final state after interacting is given by
\be
\ket{\text{final}}=i \mathcal{T}\ket{\text{initial}}= c_{kl} \ket{kl},
\ee
where $c_{kl} = i \mathcal{M}_{ij,kl}$ represents the scattering amplitude.  Note that we unit normalize the final state,  $\sum_{k,l} |c_{kl}|^2 = 1$, using an overall normalization factor.
\
Given this, the density matrix corresponding to the final state is
\begin{equation}
    \rho = \ket{\text{final}}\bra{\text{final}}
\end{equation}
and is properly normalized, $\text{Tr}(\rho)=1.$
To quantify the entanglement between the outgoing degrees of freedom, we compute the reduced density matrix $\rho_A = \text{Tr}_B(\rho)$. In components, the partial trace over subsystem $B$ reads
\begin{equation}
    (\rho_A)_{k, k'} = \sum_l \rho_{kl, k'l} .
\end{equation}
Various entanglement measures can be employed to determine the degree of entanglement of the final state. For instance, Concurrence ($C$) acts directly on the quantum state coefficients, whereas measures such as the von Neumann entropy ($S$) or the linear entropy ($S_L$) are defined via the reduced density matrix $\rho_A$. While Concurrence is particularly convenient for two-level systems (qubits), entropy-based metrics provide a framework that applies to general $d$-dimensional qudits. In particular, the linear entropy $S_L = 1 - \text{Tr}(\rho_A^2)$ is bounded by
\begin{equation}
    0 \le S_L \le 1 - \frac{1}{d} ,
\end{equation}
where $d$ denotes the dimension of the Hilbert space of subsystem $A$. Consequently, maximizing the entanglement corresponds to reaching the upper bound $S_L = 1 - 1/d$.

Notice that the scattering amplitude $\mathcal{M}$ explicitly depends on the coupling constants of the theory. Since $\rho_A$ scales quadratically with $\mathcal{M}$, computing the linear entropy generates equations of order $\mathcal{M}^4$. Determining the coupling values that maximize entanglement under these conditions becomes challenging due to the high degree of the resulting polynomials.
Alternatively, one can exploit the algebraic structure of the reduced density matrix directly. For a maximally entangled state, $\rho_A$ must be proportional to the identity matrix,
\begin{equation}
    \rho_A = \frac{1}{d} \mathbb{1} .
\end{equation}
This structure leads to the matrix element equations:
\begin{align}
    (\rho_A)_{k,k} &= \frac{1}{d}, \quad  \\
    (\rho_A)_{k,k'} &= 0 \quad (k \neq k') . 
\end{align}
The first equation indicates that the diagonal elements are equal to each other, so for maximal entanglement, we can equivalently impose the following structural constraint equations
\begin{align}
    (\rho_A)_{k,k} - (\rho_A)_{k+1,k+1}&= 0 \quad (k=1, \cdots, d-1), \nonumber \\[-0.4cm]
    & \label{eq:Structure_conditions} \\[-0.4cm]
    (\rho_A)_{k,k'} &= 0 \quad (k \neq k'). \nonumber
\end{align}
This approach reduces the problem to equations of order $\mathcal{M}^2$, which are significantly easier to solve.

In the general $N$ Higgs doublet model, the scalar sector consists of $N$ $SU(2)$ doublets of hypercharge $1/2$, $\Phi_i=(H_i^+, H_i^0 )^T$, $i=1,\dots,N$. To study the connection between quantum entanglement and the allowed $H_i H_j \to H_k H_l$ scattering processes in this model, it is necessary to define our computational basis. Following the approach in \cite{Carena:2025wyh}, we adopt the scalar flavor indices as the internal degrees of freedom, suggesting the computational basis $\{\ket{ij}\}$ defined by
\be
\{\ket{ij}\} \equiv \{\ket{H_i} \otimes \ket{H_j}\}, \quad i,j=1,\dots N.
\ee
This basis is important to the proposal of \cite{Carena:2025wyh}, which requires that the scattering of any individual computational basis states results in a maximally entangled final state.  
The bipartite states are composed of two distinguishable qudits,\footnote{Note that initial states with indistinguishable particles could be treated, as described in \cite{Chang:2024wrx}.  However, we find that those scattering channels also maximally entangle at the same parameter points.  } where the distinguishability derives from the different electroweak quantum numbers carried by the doublet components in the scattering channels, such as $H^+ H^-$ or $H^+ H^0$. While the computational basis only distinguishes the flavor labels in the initial and final states, the physically allowed states are determined by the symmetries of the potential. Consequently, we restrict our analysis to four non-vanishing scattering channels that satisfy these criteria: $H^+ H^0 \to H^+ H^0$, $H^+ H^{0^*} \to H^+ H^{0^*}$, $H^0 H^{0^*} \to H^+ H^-$, and $H^+ H^- \to H^+ H^-$.\footnote{These are non-vanishing since they conserve both charge and weak isospin.}  Other processes such as $H^0 H^{0^*} \to H^0 H^{0^*}$ and $H^+ H^- \to  H^0 H^{0^*}$ could also be considered, but their scattering matrices are equivalent to those of $H^+ H^- \to H^+ H^-$ and $H^0 H^{0^*} \to H^+ H^-$, respectively, and do not provide additional information to our entanglement analysis.

\section{2HDM}
\label{Sect:2HDM}
To illustrate the application of the formalism introduced in Section \ref{Sect:Procedure}, we first briefly examine the 2HDM case, following   \cite{Carena:2025wyh}. The most general 2HDM potential with $SU(2)\times U(1)$ symmetry is given by
\begin{align}
V_{\mathrm{2HDM}} = \ & m_{11}^2 \Phi_1^\dagger \Phi_1 + m_{22}^2 \Phi_2^\dagger \Phi_2 - \left[ m_{12}^2 \Phi_1^\dagger \Phi_2 + \text{h.c.} \right] + \frac{1}{2} \lambda_1 (\Phi_1^\dagger \Phi_1)^2 + \frac{1}{2} \lambda_2 (\Phi_2^\dagger \Phi_2)^2 \nonumber \\
& \ + \lambda_3 (\Phi_1^\dagger \Phi_1)(\Phi_2^\dagger \Phi_2) + \lambda_4 (\Phi_1^\dagger \Phi_2)(\Phi_2^\dagger \Phi_1) \nonumber \\
& \ + \left\{ \frac{1}{2} \lambda_5 (\Phi_1^\dagger \Phi_2)^2 + \left[ \lambda_6 (\Phi_1^\dagger \Phi_1) + \lambda_7 (\Phi_2^\dagger \Phi_2) \right] \Phi_1^\dagger \Phi_2 + \text{h.c.} \right\},
\end{align}
where $\Phi_i=(H_i^+, H_i^0)^T,\; i=1,2$. For simplicity, we assume $CP$-invariance so that all of the couplings $\lambda_i$ are real.  
The quartic interactions give rise to the scattering amplitudes relevant to our analysis. Explicitly, the required tree-level amplitudes $\mathcal{M}_0$ are:  
\begin{equation}
\mathcal{M}_0(H^+H^0 \to H^+H^0) = -
\begin{bmatrix}
\lambda_1 & \lambda_6 & \lambda_6 & \lambda_5 \\
\lambda_6 & \lambda_3 & \lambda_4 & \lambda_7 \\
\lambda_6 & \lambda_4 & \lambda_3 & \lambda_7 \\
\lambda_5 & \lambda_7 & \lambda_7 & \lambda_2
\end{bmatrix},
\end{equation}
\begin{equation}
\mathcal{M}_0 (H^+H^{0^*} \to H^+H^{0^*}) = -
\begin{bmatrix}
\lambda_1 & \lambda_6 & \lambda_6 & \lambda_4 \\
\lambda_6 & \lambda_3 & \lambda_5 & \lambda_7 \\
\lambda_6 & \lambda_5 & \lambda_3 & \lambda_7 \\
\lambda_4 & \lambda_7 & \lambda_7 & \lambda_2
\end{bmatrix},
\end{equation}

\begin{equation}
\mathcal{M}_0(H^0H^{0^*} \to H^+H^-) = -i 
\begin{bmatrix}
\lambda_1 & \lambda_6 & \lambda_6 & \lambda_3 \\
\lambda_6 & \lambda_4 & \lambda_5 & \lambda_7 \\
\lambda_6 & \lambda_5 & \lambda_4 & \lambda_7 \\
\lambda_3 & \lambda_7 & \lambda_7 & \lambda_2
\end{bmatrix},
\end{equation}

\begin{equation}
\text{and}\quad \mathcal{M}_0(H^+H^- \to H^+H^-) = -
\begin{bmatrix}
2\lambda_1 & 2\lambda_6 & 2\lambda_6 & \lambda_3 + \lambda_4 \\
2\lambda_6 & \lambda_3 + \lambda_4 & 2\lambda_5 & 2\lambda_7 \\
2\lambda_6 & 2\lambda_5 & \lambda_3 + \lambda_4 & 2\lambda_7 \\
\lambda_3 + \lambda_4 & 2\lambda_7 & 2\lambda_7 & 2\lambda_2
\end{bmatrix}.
\end{equation}

We wish to analyze the structure of the reduced density matrices for the final states. As an explicit example, let us consider the scattering amplitude $\mathcal{M}_0 (H^+H^0 \to H^+H^0)$ with the computational basis $\{\ket{H^+_i} \otimes \ket{H^0_j}\} \equiv \{\ket{ij}\},\ i,j=1,2$. Starting from the initial state $\ket{11}$ the scattering matrix yields the normalized final state
\be
\ket{\text{final}}_{11} = -i \ \frac{\lambda_1 \ket{11} + \lambda_6 \ket{12} + \lambda_6 \ket{21} + \lambda_5 \ket{22} }{\sqrt{\lambda_1^2 + 2\lambda_6^2 + \lambda_5^2}} .
\ee
Denoting the corresponding density matrix as $\rho^{(11)} = \ket{\text{final}}_{11} \bra{\text{final}}_{11}$, the reduced density matrix $\rho^{(11)}_A$  takes the form
\be
\rho^{(11)}_A = \frac{1}{\lambda_1^2 + 2\lambda_6^2 + \lambda_5^2}
\begin{pmatrix}
\lambda_1^2 + \lambda_6^2 & \lambda_1\lambda_6 + \lambda_5\lambda_6 \\
\lambda_1\lambda_6 + \lambda_5\lambda_6 & \lambda_5^2 + \lambda_6^2
\end{pmatrix}.
\ee
Imposing the structural conditions described in Eq.~(\ref{eq:Structure_conditions}), we obtain the equations
\begin{align}
&    \lambda_1\lambda_6 + \lambda_5\lambda_6=  0, \nonumber \\[-0.4cm] \\[-0.4cm]
 &   \dfrac{\lambda_1^2 + \lambda_6^2}{\lambda_1^2 + 2\lambda_6^2 + \lambda_5^2}= \dfrac{\lambda_5^2 + \lambda_6^2}{\lambda_1^2 + 2\lambda_6^2 + \lambda_5^2}. \nonumber
\end{align}

Applying an analogous procedure to the remaining initial states $\{\ket{12}, \ket{21}, \ket{22}\}$ leads to additional constraints on the quartic couplings. Combining these requirements across all four scattering channels and solving the resulting system, we obtain the following set of solutions:
\begin{align}
(1) &: \lambda_1 = \lambda_2 = \lambda_3 = \lambda_4 = \pm \lambda_5 \neq 0, \ \lambda_6 = \lambda_7 = 0, \label{sol:1} \\
(2) &: \lambda_1 = \lambda_2 = -\lambda_3 = -\lambda_4 =  \lambda_5 \neq 0, \ \lambda_6 = \lambda_7 = 0, \label{sol:2} \\
(3) &: \lambda_1 = -\lambda_2 = -\lambda_3 = -\lambda_4 =\pm \lambda_5 \neq 0, \ \lambda_6 = \lambda_7 = 0, \label{sol:3} \\
(4) &: \lambda_1 = -\lambda_2 = \lambda_3 = \lambda_4 = \pm \lambda_5 \neq 0, \ \lambda_6 = \lambda_7 = 0, \label{sol:4} \\
(5) &: \lambda_1 = \lambda_2 = \lambda_3 = \lambda_4 = \lambda_5 = 0, \ \lambda_6 = \lambda_7 \neq 0, \label{sol:5} \\
(6) &: \lambda_1 = \lambda_2 = -\lambda_3 = -\lambda_4 = -\lambda_5, \ \lambda_6 = -\lambda_7, \ \text{with } (\lambda_5, \lambda_7)\neq (0,0). \label{sol:6}
\end{align}
Here, each $\pm$ is a sign choice that leads to a set of solutions that are closely related.   
Solutions (1), (2), (5), and (6) in the limit $\lambda_5 = 0$ or $\lambda_7 = 0$ align with the maximal entanglement conditions reported in \cite{Carena:2025wyh}, where the Concurrence was used to quantify the entanglement of the scattering states. Although our approach reveals additional solutions for maximal entanglement (specifically, (3), (4), and (6) with $\lambda_5, \lambda_7 \neq0$), these are physically ruled out after one imposes the scalar potential is bounded-from-below.  Once this stability condition is imposed, we are left with solution (1) in Eq.~\ref{sol:1} which agrees with \cite{Carena:2025wyh}.  In particular, these coupling solutions have the extended symmetry of $U(2)^2$, which is manifest in a ``mirror" basis where the potential's quartic interactions can be written as 
\begin{equation}
V_{2HDM,sym} = \lambda \left(|\widetilde{\Phi}_1|^4+ |\widetilde{\Phi}_2|^4\right)+\cdots .\footnotemark
\end{equation}
\addtocounter{footnote}{-1}
\footnotetext{Note that the mirror basis manifestly demonstrates that requiring the basis elements to each maximally entangle is a basis dependent criteria.}
As shown in \cite{Carena:2025wyh}, this mirror structure allows the symmetries to be gauged and Yukawa interactions to be added, while maintaining maximal entanglement for Higgs scattering, as long as one keeps the interaction strengths of each sector the same.  This suggests that a generic solution of maximal entanglement occurs for mirror models where in a certain basis, one can see there are identical sectors, in terms of their particle content and self-interactions, which are decoupled from each other.  In the next section, we indeed find such solutions for the $N$HDM.   

\section{$N$ Higgs Doublet Model Max Entanglement Solutions}
\label{Sect:NHDM}
Extending to $N$ Higgs doublet models ($N$HDMs), we have found a maximally entangling solution which generalizes the mirror symmetry solutions found in \cite{Carena:2025wyh}.  For notational simplicity, we take our $N$ Higgs doublets, $\Phi_a$, where the index $a=0, \cdots, N-1$.  The mirror symmetry is manifest in the basis $\widetilde{\Phi}_a$, where 
\be
V_\text{mirror}= \lambda \sum_{a=0}^{N-1} |\widetilde{\Phi}_a|^4,
\ee
illustrating $N$ Higgs doublets with identical self-couplings that are decoupled from each other.  This is the natural generalization of the 2HDM mirror solutions in \cite{Carena:2025wyh}, where the Higgs doublets and fermions, in a certain basis, were decoupled sectors with identical self-interactions.

To go to the computational basis, $\Phi$, we define
\be
\widetilde{\Phi}_a = \frac{1}{\sqrt{N}}\sum_{b=0}^{N-1} \omega^{a b}\Phi_b, \text{ where } \omega=\exp(2\pi i/N).
\ee
Since $\omega$ is $N$-th root of unity, it can be checked that this is a unitary transformation $\widetilde{\Phi}_a= U_a^{\; b} \Phi_b$, where 
\be
U_a^{\; b} = \frac{1}{\sqrt{N}}\begin{pmatrix} 1 & 1 & 1 & \cdots \\ 1 & \omega & \omega^2 & \cdots \\ 1 & \omega^2 & \omega^4 & \cdots \\ \vdots & \vdots & \vdots & \ddots \end{pmatrix}.
\ee
In the computational basis, the potential can be shown to be
\be
V_\text{mirror}= \frac{\lambda}{N} \sum_{a=0}^{N-1} \left|\sum_{b=0}^{N-1} \Phi_b^\dag \Phi_{b+a}\right|^2.
\ee
Note that in this form, the potential exhibits a $Z_N$ symmetry, where $\Phi_a \to z^a \Phi_a$. 
One can now calculate the scattering matrix for a neutral Higgs with a charged Higgs, giving 
\be
M(\Phi^0_a \Phi^+_b \to \Phi^0_c \Phi^+_d) = \frac{2\lambda}{N} \delta_{a+b, c+d}. 
\ee
The Kronecker delta terms are defined modulo $N$, i.e.~$\delta_{a,b} =1$ only if $a-b=0 \text{ mod } N$, consistent with the $Z_N$ charges.  For any choice of $a,b$, there are always $N$ possible solutions for $c,d.$  In this way, this potential scatters each incoming product state (of the computational basis) into a linear combination of $N$ product states with equal coefficients, leading to maximal entanglement.  Similarly, the other scattering channels have the same type of flavor structure and thus will  also maximally entangle. 

In \cite{Carena:2025wyh}, for the two Higgs doublet model, it was also shown that it was possible to gauge $U(2)^2$ and add Yukawa couplings of the Higgs doublets to fermions in a way that maintained maximal entanglement.  In the the analogous $\widetilde{\Phi}_1, \widetilde{\Phi}_2$ basis, this  requires the $\widetilde{\Phi}_a$ to be a doublet under $U(2)_a$ where the gauge couplings of the two gauge groups are equal.  Similarly, the fermion content is mirrored with $\widetilde{\Phi}_a$ coupling to a set of fermions $\Psi_a$ with identical Yukawa couplings.  Transforming back to the computational basis $\Phi_a$, the scattering amplitudes then have the correct maximal entangling behavior.  This mirror structure can similarly be extended to the $N$ Higgs doublet solution, where in the $\widetilde{\Phi}$ basis, we have $N$ decoupled sectors, each sector containing an identical set of interactions amongst the Higgs bosons, gauge bosons, and fermions.  To summarize, mirror models for the $N$HDM naturally have maximal entanglement when written in the computational basis.  However, there may be other solutions without extended symmetries.  To explore this possibility, we now look in detail at the three Higgs doublet model.  
\newpage
\section{3HDM}
\label{Sect:3HDM}
Working with massless fields, the most general 3HDM potential with $SU(2)\times U(1)$ symmetry is given by
\begin{align}
V_{\mathrm{3HDM}} = 
 &
\ \lambda_{11} (\Phi_1^{\dagger} \Phi_1)^2 
+ \lambda_{22} (\Phi_2^{\dagger} \Phi_2)^2 
+ \lambda_{33} (\Phi_3^{\dagger} \Phi_3)^2
     \nonumber \\ &
+ \lambda_{1122} (\Phi_1^{\dagger} \Phi_1) (\Phi_2^{\dagger} \Phi_2) 
+ \lambda_{1133} (\Phi_1^{\dagger} \Phi_1) (\Phi_3^{\dagger} \Phi_3) 
+ \lambda_{2233} (\Phi_2^{\dagger} \Phi_2) (\Phi_3^{\dagger} \Phi_3) 
    \nonumber \\ &
+ \lambda_{1221} (\Phi_1^{\dagger} \Phi_2) (\Phi_2^{\dagger} \Phi_1)
+ \lambda_{1331} (\Phi_1^{\dagger} \Phi_3) (\Phi_3^{\dagger} \Phi_1) 
+ \lambda_{2332} (\Phi_2^{\dagger} \Phi_3) (\Phi_3^{\dagger} \Phi_2)
     \nonumber \\ &
 + \bigg[\, {\lambda_{1212} \over 2} (\Phi_1^{\dagger} \Phi_2)^2 
+ {\lambda_{1313} \over 2} (\Phi_1^{\dagger} \Phi_3)^2 
+ {\lambda_{2323} \over 2}  (\Phi_2^{\dagger} \Phi_3)^2 
\nonumber \\ & 
+ \lambda_{1213}  (\Phi_1^{\dagger} \Phi_2) (\Phi_1^{\dagger} \Phi_3)
+ \lambda_{2113} (\Phi_2^{\dagger} \Phi_1) (\Phi_1^{\dagger} \Phi_3)
+ \lambda_{1323}    (\Phi_1^{\dagger} \Phi_3) (\Phi_2^{\dagger} \Phi_3)
\label{eq:V3HDM} \\ & 
+ \lambda_{1332}   (\Phi_1^{\dagger} \Phi_3) (\Phi_3^{\dagger} \Phi_2)
+ \lambda_{2123}  (\Phi_2^{\dagger} \Phi_1) (\Phi_2^{\dagger} \Phi_3)
+ \lambda_{1223} (\Phi_1^{\dagger} \Phi_2) (\Phi_2^{\dagger} \Phi_3) 
     \nonumber \\ & 
+ \lambda_{1112}  (\Phi_1^{\dagger} \Phi_1) (\Phi_1^{\dagger} \Phi_2)
+ \lambda_{2212}  (\Phi_2^{\dagger} \Phi_2) (\Phi_1^{\dagger} \Phi_2)  
+ \lambda_{1113}  (\Phi_1^{\dagger} \Phi_1) (\Phi_1^{\dagger} \Phi_3) 
     \nonumber \\ & 
+ \lambda_{1123}  (\Phi_1^{\dagger} \Phi_1) (\Phi_2^{\dagger} \Phi_3)
+ \lambda_{2213}  (\Phi_2^{\dagger} \Phi_2) (\Phi_1^{\dagger} \Phi_3)  
+ \lambda_{2223}  (\Phi_2^{\dagger} \Phi_2) (\Phi_2^{\dagger} \Phi_3) 
     \nonumber \\ & 
+ \lambda_{3312}  (\Phi_3^{\dagger} \Phi_3) (\Phi_1^{\dagger} \Phi_2) 
+ \lambda_{3313}  (\Phi_3^{\dagger} \Phi_3)(\Phi_1^{\dagger} \Phi_3) 
+ \lambda_{3323}  (\Phi_3^{\dagger} \Phi_3) (\Phi_2^{\dagger} \Phi_3) 
+{\mathrm{h.c.}}\bigg],\; \nonumber
\end{align}
where $\Phi_i=(H_i^+, H_i^0)^T,\; i=1,2,3$, are the Higgs doublets. We again assume CP conservation, so the coupling constants are real. 

Proceeding with the calculation of the relevant tree-level scattering matrices, we obtain the following results:

\begin{equation}
\mathcal{M}_0(H^+ H^0 \to H^+ H^0) = - \begin{bmatrix}
2\lambda_{11} & \lambda_{1112} & \lambda_{1113} & \lambda_{1112} & \lambda_{1212} & \lambda_{1213} & \lambda_{1113} & \lambda_{1213} & \lambda_{1313} \\
\lambda_{1112} & \lambda_{1122} & \lambda_{1123} & \lambda_{1221} & \lambda_{2212} & \lambda_{1223} & \lambda_{2113} & \lambda_{2213} & \lambda_{1323} \\
\lambda_{1113} & \lambda_{1123} & \lambda_{1133} & \lambda_{2113} & \lambda_{2123} & \lambda_{3312} & \lambda_{1331} & \lambda_{1332} & \lambda_{3313} \\
\lambda_{1112} & \lambda_{1221} & \lambda_{2113} & \lambda_{1122} & \lambda_{2212} & \lambda_{2213} & \lambda_{1123} & \lambda_{1223} & \lambda_{1323} \\
\lambda_{1212} & \lambda_{2212} & \lambda_{2123} & \lambda_{2212} & 2\lambda_{22} & \lambda_{2223} & \lambda_{2123} & \lambda_{2223} & \lambda_{2323} \\
\lambda_{1213} & \lambda_{1223} & \lambda_{3312} & \lambda_{2213} & \lambda_{2223} & \lambda_{2233} & \lambda_{1332} & \lambda_{2332} & \lambda_{3323} \\
\lambda_{1113} & \lambda_{2113} & \lambda_{1331} & \lambda_{1123} & \lambda_{2123} & \lambda_{1332} & \lambda_{1133} & \lambda_{3312} & \lambda_{3313} \\
\lambda_{1213} & \lambda_{2213} & \lambda_{1332} & \lambda_{1223} & \lambda_{2223} & \lambda_{2332} & \lambda_{3312} & \lambda_{2233} & \lambda_{3323} \\
\lambda_{1313} & \lambda_{1323} & \lambda_{3313} & \lambda_{1323} & \lambda_{2323} & \lambda_{3323} & \lambda_{3313} & \lambda_{3323} & 2\lambda_{33}
\end{bmatrix},
\end{equation}
\begin{equation}
\mathcal{M}_0(H^+ H^{0^*} \to H^+ H^{0^*}) = - 
\begin{bmatrix}
2\lambda_{11} & \lambda_{1112} & \lambda_{1113} & \lambda_{1112} & \lambda_{1221} & \lambda_{2113} & \lambda_{1113} & \lambda_{2113} & \lambda_{1331} \\
\lambda_{1112} & \lambda_{1122} & \lambda_{1123} & \lambda_{1212} & \lambda_{2212} & \lambda_{2123} & \lambda_{1213} & \lambda_{2213} & \lambda_{1332} \\
\lambda_{1113} & \lambda_{1123} & \lambda_{1133} & \lambda_{1213} & \lambda_{1223} & \lambda_{3312} & \lambda_{1313} & \lambda_{1323} & \lambda_{3313} \\
\lambda_{1112} & \lambda_{1212} & \lambda_{1213} & \lambda_{1122} & \lambda_{2212} & \lambda_{2213} & \lambda_{1123} & \lambda_{2123} & \lambda_{1332} \\
\lambda_{1221} & \lambda_{2212} & \lambda_{1223} & \lambda_{2212} & 2\lambda_{22} & \lambda_{2223} & \lambda_{1223} & \lambda_{2223} & \lambda_{2332} \\
\lambda_{2113} & \lambda_{2123} & \lambda_{3312} & \lambda_{2213} & \lambda_{2223} & \lambda_{2233} & \lambda_{1323} & \lambda_{2323} & \lambda_{3323} \\
\lambda_{1113} & \lambda_{1213} & \lambda_{1313} & \lambda_{1123} & \lambda_{1223} & \lambda_{1323} & \lambda_{1133} & \lambda_{3312} & \lambda_{3313} \\
\lambda_{2113} & \lambda_{2213} & \lambda_{1323} & \lambda_{2123} & \lambda_{2223} & \lambda_{2323} & \lambda_{3312} & \lambda_{2233} & \lambda_{3323} \\
\lambda_{1331} & \lambda_{1332} & \lambda_{3313} & \lambda_{1332} & \lambda_{2332} & \lambda_{3323} & \lambda_{3313} & \lambda_{3323} & 2\lambda_{33}
\end{bmatrix},
\end{equation}
\begin{equation}
\mathcal{M}_0(H^0 H^{0^*} \to H^+ H^-) = -
\begin{bmatrix}
2\lambda_{11} & \lambda_{1112} & \lambda_{1113} & \lambda_{1112} & \lambda_{1122} & \lambda_{1123} & \lambda_{1113} & \lambda_{1123} & \lambda_{1133} \\
\lambda_{1112} & \lambda_{1221} & \lambda_{2113} & \lambda_{1212} & \lambda_{2212} & \lambda_{2123} & \lambda_{1213} & \lambda_{1223} & \lambda_{3312} \\
\lambda_{1113} & \lambda_{2113} & \lambda_{1331} & \lambda_{1213} & \lambda_{2213} & \lambda_{1332} & \lambda_{1313} & \lambda_{1323} & \lambda_{3313} \\
\lambda_{1112} & \lambda_{1212} & \lambda_{1213} & \lambda_{1221} & \lambda_{2212} & \lambda_{1223} & \lambda_{2113} & \lambda_{2123} & \lambda_{3312} \\
\lambda_{1122} & \lambda_{2212} & \lambda_{2213} & \lambda_{2212} & 2\lambda_{22} & \lambda_{2223} & \lambda_{2213} & \lambda_{2223} & \lambda_{2233} \\
\lambda_{1123} & \lambda_{2123} & \lambda_{1332} & \lambda_{1223} & \lambda_{2223} & \lambda_{2332} & \lambda_{1323} & \lambda_{2323} & \lambda_{3323} \\
\lambda_{1113} & \lambda_{1213} & \lambda_{1313} & \lambda_{2113} & \lambda_{2213} & \lambda_{1323} & \lambda_{1331} & \lambda_{1332} & \lambda_{3313} \\
\lambda_{1123} & \lambda_{1223} & \lambda_{1323} & \lambda_{2123} & \lambda_{2223} & \lambda_{2323} & \lambda_{1332} & \lambda_{2332} & \lambda_{3323} \\
\lambda_{1133} & \lambda_{3312} & \lambda_{3313} & \lambda_{3312} & \lambda_{2233} & \lambda_{3323} & \lambda_{3313} & \lambda_{3323} & 2\lambda_{33}
\end{bmatrix},
\end{equation}
\begin{align}
\text{and}\quad & \mathcal{M}_0 (H^+ H^- \to H^+ H^-) = \\
&-
\resizebox{0.9\textwidth}{!}{$%
\begin{bmatrix}
4\lambda_{11} & 2\lambda_{1112} & 2\lambda_{1113} & 2\lambda_{1112} & \lambda_{1122} + \lambda_{1221} & \lambda_{1123} + \lambda_{2113} & 2\lambda_{1113} & \lambda_{1123} + \lambda_{2113} & \lambda_{1133} + \lambda_{1331} \\
2\lambda_{1112} & \lambda_{1122} + \lambda_{1221} & \lambda_{1123} + \lambda_{2113} & 2\lambda_{1212} & 2\lambda_{2212} & 2\lambda_{2123} & 2\lambda_{1213} & \lambda_{1223} + \lambda_{2213} & \lambda_{1332} + \lambda_{3312} \\
2\lambda_{1113} & \lambda_{1123} + \lambda_{2113} & \lambda_{1133} + \lambda_{1331} & 2\lambda_{1213} & \lambda_{1223} + \lambda_{2213} & \lambda_{1332} + \lambda_{3312} & 2\lambda_{1313} & 2\lambda_{1323} & 2\lambda_{3313} \\
2\lambda_{1112} & 2\lambda_{1212} & 2\lambda_{1213} & \lambda_{1122} + \lambda_{1221} & 2\lambda_{2212} & \lambda_{1223} + \lambda_{2213} & \lambda_{1123} + \lambda_{2113} & 2\lambda_{2123} & \lambda_{1332} + \lambda_{3312} \\
\lambda_{1122} + \lambda_{1221} & 2\lambda_{2212} & \lambda_{1223} + \lambda_{2213} & 2\lambda_{2212} & 4\lambda_{22} & 2\lambda_{2223} & \lambda_{1223} + \lambda_{2213} & 2\lambda_{2223} & \lambda_{2233} + \lambda_{2332} \\
\lambda_{1123} + \lambda_{2113} & 2\lambda_{2123} & \lambda_{1332} + \lambda_{3312} & \lambda_{1223} + \lambda_{2213} & 2\lambda_{2223} & \lambda_{2233} + \lambda_{2332} & 2\lambda_{1323} & 2\lambda_{2323} & 2\lambda_{3323} \\
2\lambda_{1113} & 2\lambda_{1213} & 2\lambda_{1313} & \lambda_{1123} + \lambda_{2113} & \lambda_{1223} + \lambda_{2213} & 2\lambda_{1323} & \lambda_{1133} + \lambda_{1331} & \lambda_{1332} + \lambda_{3312} & 2\lambda_{3313} \\
\lambda_{1123} + \lambda_{2113} & \lambda_{1223} + \lambda_{2213} & 2\lambda_{1323} & 2\lambda_{2123} & 2\lambda_{2223} & 2\lambda_{2323} & \lambda_{1332} + \lambda_{3312} & \lambda_{2233} + \lambda_{2332} & 2\lambda_{3323} \\
\lambda_{1133} + \lambda_{1331} & \lambda_{1332} + \lambda_{3312} & 2\lambda_{3313} & \lambda_{1332} + \lambda_{3312} & \lambda_{2233} + \lambda_{2332} & 2\lambda_{3323} & 2\lambda_{3313} & 2\lambda_{3323} & 4\lambda_{33}
\end{bmatrix}.$ \nonumber %
}
\end{align}
\newline
With these matrices in hand, we now determine the relations among the $\lambda$ couplings that produce maximal entanglement, following the procedure outlined in Section \ref{Sect:Procedure}. 

Compared to a 2HDM, in a 3HDM there are significantly more equations obtained from maximizing entanglement in the final scattering states. Actually, we can try to reduce the number of equations by using the fact that the scattering amplitudes of our different scattering processes, and consequently the reduced density matrices, are symmetric. Thus, the second condition from Eq.~ (\ref{eq:Structure_conditions}) can be applied to the upper-triangle elements of the reduced density matrix, ignoring the lower-triangle. Following our framework, in the $N$ Higgs doublet model, each scattering process has $N^2$ initial possible states, and the reduced density matrix of each corresponding final state provides $\frac{N^2+N-2}{2}$ equations. For $N=2$ it is possible to find analytical solutions, but for $N=3$ it becomes challenging. Instead, we chose to analyze this numerically to characterize solutions and then followed them up, to get an analytic understanding.

Let $\{E_i = 0\}$ denote the set of equations obtained by implementing Eq.~(\ref{eq:Structure_conditions}) on the reduced density matrices of all final states within our computational basis, considering the four scattering processes analyzed. If we define
\be
 P=\sum_i E_i^2,
\ee
the minimum of $P$ is $0$ by construction, and corresponds to  solving all the equations $E_i = 0$ simultaneously. Since we expect to find relations among the coupling constants, it is useful to define a unit vector $\hat{\lambda}$ composed of these couplings, such that $||\hat{\lambda}||^2 = \lambda_{ii}^2 + \lambda_{ijkl}^2 = 1$.  Essentially, the idea is to search for the specific directions of $\hat{\lambda}$ that minimize $P(\hat{\lambda})$, where the parameter space can be explored numerically by randomly initializing the coupling values to serve as starting points for the minimizer. In practice, this numerical approach is implemented by minimizing
\be
Q = \sum_i \left(E_i \right)^2 + \left(||\hat{\lambda}||^2 - 1 \right)^2 .
\ee

We executed this method in \textit{Mathematica} using $1000$ independent random initializations to exhaustively explore the structure of the solutions. Constraining the couplings to the unit hypersphere helps to eliminate redundant scaling solutions where the configurations differ only by an overall multiplicative factor. Furthermore, sorting the numerical solutions allows us to easily distinguish and classify the distinct solution classes. Since the expressions $E_i$ are quadratic in $\lambda_{ijkl}$, we can anticipate that many solutions differ only by sign combinations. Although the numerical method described above efficiently distinguishes the proportional relationship among the couplings within each solution class, generating all allowed sign variations purely numerically would require a significantly larger number of runs. Therefore, once the overall solution structure is identified numerically, we complement the process by analytically determining the permitted sign configurations.  Note that as a check of this numerical procedure, we also confirmed that we could find all of the solutions of the 2HDM in this way.  

As an example in the 3HDM, following the numerical scan, we could determine that one family of solutions satisfies
\begin{align}
&    |\lambda_{2233}| = |\lambda_{2332}| = |\lambda_{1213}| = |\lambda_{2323}| = |\lambda_{2113}| = |\lambda_{2212}| = |\lambda_{1123}| = |\lambda_{3313}| = 2|\lambda_{11}|, \nonumber \\  
 &   \lambda_{ii} = \lambda_{ijkl} =0,\quad \text{otherwise.} 
\end{align}
The numerical sampling also revealed several sign variations among the $\lambda$ parameters, although not necessarily all of them. The allowed sign choices were determined analytically in accordance with the equations $\{E_i = 0\}$, and as a result, the constraints $\lambda_{2233}=\lambda_{2332}$ and $\lambda_{2113}=\lambda_{1123}$ were found. Beyond these, no further constraints apply, and the signs of the parameters remain free.

Our analysis yields six distinct families of solutions, which we describe below. Among the obtained solutions, some are equivalent under an overall sign flip transformation, so that, if $\{\lambda_{ii}, \lambda_{ijkl}\}$ forms a valid parameter set, so is $\{-\lambda_{ii}, -\lambda_{ijkl}\}$. Similarly, other solutions are related by field redefinitions of the form $\Phi_i \to -\Phi_i$. Since these transformations do not lead to physically distinct configurations, we adopt the convention where $\lambda_{11} = +\lambda $. Additionally, to eliminate the residual redundancy associated with the field redefinitions $\Phi_1 \to -\Phi_1$ and $\Phi_2 \to -\Phi_2$,\footnote{Since each term in the potential involves four of the $\Phi_1, \Phi_2, \Phi_3$'s, the transformation where all three Higgs doublets flip sign, $\Phi_i\to -\Phi_i$, leaves the potential unchanged.  }  in each class of solution we fix two parameters containing an odd number of indices $1$ and $2$. As a result of these conventions, the families of solutions are given by

\begin{align}
    (1): \qquad
    & \lambda_{11} = \lambda, 
    \quad \lambda_{2233} = \lambda_{2332} = \pm 2\lambda, \quad \lambda_{1213} = \pm 2\lambda, 
    \quad \lambda_{2323} = \pm 2\lambda, 
    \quad \lambda_{2212} = 2\lambda, 
    \nonumber \\[-0.4cm]
    & \\[-0.4cm]
   & \lambda_{2113} = \lambda_{1123} = 2\lambda, 
    \quad \lambda_{3313} = \pm 2\lambda,  
    \quad \lambda_{ii} = \lambda_{ijkl} =0 \quad \text{otherwise.} \nonumber \\[10pt]
    (2): \qquad
    & \lambda_{11} = \lambda, 
    \quad \lambda_{2233} = \lambda_{2332} = \pm 2\lambda, \quad \lambda_{1213} = 2\lambda, 
    \quad \lambda_{22} = \pm \lambda, 
    \quad \lambda_{1133} = \lambda_{1331} = \pm 2\lambda,  \nonumber \\
    &  \lambda_{2123} = 2\lambda, 
    \quad \lambda_{33} = \pm \lambda,  
    \quad \lambda_{1122} = \lambda_{1221} = \pm 2\lambda,
    \quad \lambda_{1323} = \pm 2\lambda,  \\
    & \lambda_{ii} = \lambda_{ijkl} = 0 \quad \text{otherwise.} \nonumber \\[10pt]
    (3): \qquad
    & 4 \lambda = 4\lambda_{11} = 4\lambda_{22} =  4\lambda_{33} = 2\lambda_{2233} = 2\lambda_{1133} = 2\lambda_{1122} = 2\lambda_{2332} = 2\lambda_{1331} = 2\lambda_{1221} \nonumber \\ 
    &\quad = -\lambda_{2323} = -\lambda_{1313} = -\lambda_{1212} = \lambda_{2113} = \lambda_{1123} = \lambda_{3323} = \lambda_{2223} = -2\lambda_{1213} \nonumber \\[-0.4cm] \\[-0.4cm]
    &\quad = \lambda_{3313} = \lambda_{1223} = \lambda_{2213} = \lambda_{1113} = -2\lambda_{2123} = -\lambda_{2212} = -\lambda_{1112} = -\lambda_{1332}  \nonumber \\
    &\quad = -\lambda_{3312} = 2\lambda_{1323}. \nonumber \\[10pt]
    (4): \qquad
&    \lambda_{2233} = \lambda_{2332} = \lambda_{2323} = \lambda_{1213} = \lambda_{2113} = \lambda_{1123} = \lambda_{3313} = \lambda_{2212} = 2 \lambda_{11} = 2\lambda_{1}, \nonumber \\
 &   \lambda_{1133} = \lambda_{1331} = \lambda_{1313} = \lambda_{2123} = \lambda_{1223} = \lambda_{2213} = \lambda_{3323} = \lambda_{1112} = 2 \lambda_{22} = 2\lambda_{2},  \\
 &   \lambda_{1122} = \lambda_{1221} = \lambda_{1212} = \lambda_{2223} = \lambda_{1113} = \lambda_{1323} = \lambda_{1332} = \lambda_{3312} = 2 \lambda_{33} = 2\lambda_{3}. \nonumber
    \end{align}
In Solution (4),  the  parameters $\lambda_1$, $\lambda_2$ and $\lambda_3$ satisfy the constraint 
\begin{equation}
\frac{1}{\lambda_1} + \frac{1}{\lambda_2} + \frac{1}{\lambda_3} = 0.
\end{equation}
In each of these solution families, the $\pm$ indicates an independent sign choice.  Thus, Solution (1) really represents $2^4=16$ types of solutions and Solution (2) represents $2^6=64$ types of solutions.   There are also two remaining families of solutions that are not listed.  They can be generated directly from Solution (1) through the cyclic index permutations $(1 \to 2, 2 \to 3, 3 \to 1)$ and $(1 \to 3, 2 \to 1, 3 \to 2)$. Since these solutions are physically equivalent, we continue the analysis using Solution (1) as a representative case.

\subsection{Analysis of solutions}
We have listed four families of solutions whose resulting potential generates maximally entangled states. However, not all of these solutions produce potentials bounded from below. This feature can be checked systematically using a numerical approach, which can then be followed up analytically to find unstable directions or prove stability.  Again, we checked that this method agreed with the earlier 2HDM results. Expressed in terms of the real fields $\phi_i$, the scalar potential $V_{\mathrm{3HDM}}(\vec{x})$ with $\vec{x}=(\phi_1,\phi_2,...\phi_{12})$ is a homogeneous polynomial of degree 4. Due to this homogeneity, positivity over the entire field space is equivalent to positivity restricted to the unit hypersphere $||\vec{x}||=1$. Therefore, the potential is stable if and only if the global minimum on the hypersphere satisfies
\be
\text{min}_{||\vec{x}||=1} V_{\mathrm{3HDM}}(\vec{x})>0. \label{eq:Stability_num}
\ee
Analytically, we can identify and rule out solutions that produce unstable scalar potentials by evaluating the potential along specific field directions. In general, for any case where $\lambda_{ii}\neq 0$, a necessary condition for the potential to be bounded from below is $\lambda_{ii}>0$, which can be shown by considering the field directions $\Phi_i=(0,\phi_i)^T$ and $\Phi_j=0 \ (j\neq i)$. Additionally, it can be determined that all solutions belonging to Solution (1), (3) and (4) are unstable, as discussed below:

For Solution (1), taking the direction where $\Phi_1=(0,\phi_1)^T$, $\Phi_2=(0,\phi_2)^T$ and $\Phi_3=0$, the potential reduces to 
\be
V_{\mathrm{3HDM}} = \lambda(\phi_1 ^4 + 4 \phi_1 \phi_2^3 ),
\ee
where $\lambda>0$. The sign of $\phi_1$ can always be chosen so that the second term is negative. Then, in the limit $\phi_2\to \infty$ the potential diverges to $-\infty$, proving that all potentials generated by Solution (1) are unstable.

In the case of Solution (3), projecting along the field directions where $\Phi_1 = (0,\phi)^T$, $\Phi_2 = (0,\phi)^T$ and $\Phi_3 = (0,-\phi)^T$, the potential takes the form
\be
V_{\mathrm{3HDM}} = -81 \lambda \phi^4,
\ee
indicating that it is an unstable potential.

Solution (4) is subject to the restriction 
\begin{equation}
\frac{1}{\lambda_1} + \frac{1}{\lambda_2} + \frac{1}{\lambda_3} = 0. \nonumber
\end{equation}
For this equality to hold, at least one coupling $\lambda_i$ must be strictly negative. As a result, the necessary condition $\lambda_{ii} > 0$ is violated, and Solution (4) is ruled out. 

This leaves the solutions in Solution (2) as the only possible potentials which maximize entanglement and are bounded from below. Following the same strategy, we can begin by determining the signs of $\lambda_{iijj}$. Taking the field directions $\Phi_i=(0,\phi_i)^T$, $\Phi_j=(0,\phi_j)^T$ and $\Phi_k=0$ ($i\neq j\neq k$), the resulting potential simplifies to
\be
V_{\mathrm{3HDM}} =\lambda(\phi_i ^4 + \phi_j^4 \pm 4  \phi_i^2 \phi_j^2),
\ee
where $\lambda>0$ and the $\pm$ sign arises from $\lambda_{iijj}= \pm 2\lambda$. Clearly, this sign must be positive for the potential to be bounded from below, yielding $\lambda_{iijj} = 2 \lambda$. The remaining solutions in this class to be evaluated are
\begin{align}
\lambda_{2233} &= \lambda_{2332} =  \lambda_{1213} = \lambda_{1133} = \lambda_{1331} = \lambda_{2123} = \lambda_{1122} \nonumber \\
 &= \lambda_{1221} = \pm \lambda_{1323} = 2 \lambda_{11} = 2 \lambda_{22} = 2 \lambda_{33} = 2\lambda>0, \nonumber \\
 \lambda_{ijkl} &= 0 \quad \text{for all other } i,j,k,l.
\end{align}
The scalar potentials corresponding to the solutions with $\lambda_{1323} = \pm 2\lambda$ can be written as


\begin{align}
V_{\mathrm{3HDM}}^{(+)}& = \lambda \left[ (\Phi^\dagger \Phi)^2 + 2 \left| \Phi^\dagger S^{+} \Phi \right|^2 \right], \ \text{and} \\
V_{\text{3HDM}}^{(-)} &=
\lambda \Big[
\Big(\sum_{i=1}^3|\Phi_i|^2 \Big)^2
+ 2\Big(
 |\Phi_1^\dagger \Phi_2|^2
+ |\Phi_2^\dagger \Phi_3|^2
+ |\Phi_3^\dagger \Phi_1|^2  \nonumber \\
&\quad + \left\{ (\Phi_1^\dagger \Phi_2)(\Phi_1^\dagger \Phi_3) + (\Phi_2^\dagger \Phi_3)(\Phi_2^\dagger \Phi_1) - (\Phi_3^\dagger \Phi_1)(\Phi_3^\dagger \Phi_2) + {\mathrm{h.c.}} \right\}\Big)
\Big] \label{eq:V---(1)} \\
&= \lambda \left[ (\Phi^\dagger \Phi)^2 + 2 \sum_{a=1}^3 \left| \Phi^\dagger S_a^- \Phi \right|^2 - 2 \sum_{a=4}^6 \left| \Phi^\dagger S_a^- \Phi \right|^2 \right] \label{eq:V---(2)},
\end{align}

where $\Phi=(\Phi_1, \Phi_2, \Phi_3)^T$ and 
\begin{align}
    S^+ &= \begin{pmatrix}
    0 & 1 & 0 \\
    0 & 0 & 1 \\
    1 & 0 & 0
    \end{pmatrix}, \quad
    S_1^- = \begin{pmatrix}
    0 & 1 & 0 \\
    0 & 0 & 0 \\
    1 & 0 & 0
    \end{pmatrix}, 
    \quad
    S_2^- = \begin{pmatrix}
    0 & 1 & 0 \\
    0 & 0 & 1 \\
    0 & 0 & 0
    \end{pmatrix}, 
    \quad
    S_3^- = \begin{pmatrix}
    0 & 0 & 0 \\
    0 & 0 & -1 \\
    1 & 0 & 0
    \end{pmatrix}, \nonumber \\
    S_4^- &= \begin{pmatrix}
    0 & 1 & 0 \\
    0 & 0 & 0 \\
    0 & 0 & 0
    \end{pmatrix}, 
    \quad
    S_5^- = \begin{pmatrix}
    0 & 0 & 0 \\
    0 & 0 & 1 \\
    0 & 0 & 0
    \end{pmatrix}, 
    \quad
    S_6^- = \begin{pmatrix}
    0 & 0 & 0 \\
    0 & 0 & 0 \\
    1 & 0 & 0
    \end{pmatrix}. \label{eqn:smatrices}
\end{align}
The sum of squared scalar bilinears in $V_{\mathrm{3HDM}}^{(+)}$ ensures that the potential is bounded from below. On the other hand, stability is not immediately obvious for $V_{\mathrm{3HDM}}^{(-)}$ due to the relative negative signs. 
By expanding $V_{\mathrm{3HDM}}^{(-)}$, we can write it in an alternative form which is manifestly positive definite
\begin{align}
V_{\text{3HDM}}^{(-)} &=
\lambda \Big[
\Big(\sum_{i=1}^3|\Phi_i|^4 \Big)\\
& + 2\Big(\left( |\Phi_1|^2 |\Phi_2|^2- |\Phi_1^\dag \Phi_2|^2\right)+ 
\left( |\Phi_1|^2 |\Phi_3|^2- |\Phi_1^\dag \Phi_3|^2\right) + \left( |\Phi_2|^2 |\Phi_3|^2- |\Phi_2^\dag \Phi_3|^2 \right)  \nonumber \\
& \quad \quad +|\Phi^\dag_1 \Phi_2 + \Phi^\dag_3 \Phi_1|^2 +
|\Phi^\dag_3 \Phi_2 + \Phi^\dag_2 \Phi_1|^2+|\Phi^\dag_3 \Phi_1 - \Phi^\dag_2 \Phi_3|^2\Big) \Big].
\end{align}
By the Cauchy-Schwarz inequality, the terms on the second line are positive definite, so the whole expression is then non-negative.  In fact, a detailed analysis shows that the minimum of the potential on the unit hypersphere is $\lambda/3.$

So far we have shown that $V_{\text{3HDM}}^{(+)}$ and $V_{\text{3HDM}}^{(-)}$ are stable scalar potentials, now let us proceed with the discussion of their symmetries, using the form of Eq.~ (\ref{eq:V---(2)}) to describe $V_{\text{3HDM}}^{(-)}$. The first term in both potentials $(\Phi^\dagger \Phi)^2$ is invariant under $U(6)$, so we can begin by looking the subgroup of $U(6)$ under which the remaining terms are also invariant. The generators we consider for the symmetry subgroup are given by the product group set 
\be
\{T^A \otimes \mathbb{1}, \quad T^A \otimes \dfrac{\sigma^a}{2}\}
\ee
where $T^A$ denotes the generators acting on the  3-dimensional flavor space, $\sigma^a$ $(a=1,2,3)$ are the Pauli matrices acting on the weak isospin space, and $ \mathbb{1}$ is the $2\times2$ identity matrix representing weak hypercharge. For a quartic operator of the form $|\Phi^\dagger S \Phi|^2$ to be invariant under the symmetry with generator $T^A$ in the flavor component, it is required that
\be
\left[ S,T^A\right]=0 \quad \text{and} \quad \left[ S^\dagger,T^A\right]=0.\label{eq:Generator_relation}
\ee
Note that if there is more than one generator, then the generic combination $\alpha_A T^A$ satisfies the relation, where the sum over the index $A$ is implicit.

Starting with $V_{\mathrm{3HDM}}^{(+)}$, the operator $\alpha_A T^A$ that satisfies the relation from Eq.~(\ref{eq:Generator_relation}) with $S = S^+$ (see Eq.~(\ref{eqn:smatrices})), is 
\begin{align}
\alpha_A T^A &=  \begin{pmatrix}
    \alpha_0 & \alpha_2 +i \alpha_3 & \alpha_2 - i \alpha_3 \\
    \alpha_2 - i \alpha_3 & \alpha_0 & \alpha_2 + i \alpha_3 \\
    \alpha_2 + i \alpha_3 & \alpha_2 - i \alpha_3 & \alpha_0
    \end{pmatrix} .
    \label{eq:alpha T}
\end{align}
This determines the generators
\be
T^0 = \begin{pmatrix}
    1 & 0 & 0 \\
    0 & 1 & 0 \\
    0 & 0 & 1
    \end{pmatrix}, \quad T^1 = \begin{pmatrix}
    0 & 1 & 1 \\
    1 & 0 & 1 \\
    1 & 1 & 0
    \end{pmatrix} \quad \text{and} \quad T^2 = \begin{pmatrix}
    0 & i & -i \\
    -i & 0 & i \\
    i & -i & 0
    \end{pmatrix}.
\ee
As a result, the symmetry group of $V_{\mathrm{3HDM}}^{(+)}$ has a total of $12$ generators. In addition, one can verify that the commutation relation $[T^A,T^B]=0$ holds, and therefore the symmetry of the potential is $U(2) \times U(2) \times U(2)=U(2)^3$. This represents a symmetry enhancement beyond the minimal $U(2)$ of the general potential.

The expression in Eq.~(\ref{eq:alpha T}) can be diagonalized using the unitary matrix
\be
U=\frac{1}{\sqrt{3}}\begin{pmatrix}
    1 & 1 & 1 \\
    1 & \omega & \omega^2 \\
    1 & \omega^2 & \omega^4
    \end{pmatrix}
\ee
through $\alpha_A T^A = U^\dagger D U$, where D is the diagonal matrix and $\omega=e^{2\pi i /3}$. This suggests the field redefinitions
\begin{align}
    \widetilde{\Phi}_1 &= \frac{1}{\sqrt{3}}(\Phi_1+\Phi_2+\Phi_3) \nonumber \\
    \widetilde{\Phi}_2 &= \frac{1}{\sqrt{3}}(\Phi_1 + \omega \Phi_2 + \omega^2 \Phi_3) \\
    \widetilde{\Phi}_3 &= \frac{1}{\sqrt{3}}(\Phi_1 + \omega^2 \Phi_2 + \omega^4 \Phi_3) \nonumber 
\end{align}
under which the potential reads
\be
V_{\mathrm{3HDM}}^{(+)}= \lambda \sum_{i=1}^3 |\widetilde{\Phi}_i|^4,
\ee
showing the three decoupled sectors $\{ \widetilde{\Phi}_1, \widetilde{\Phi}_2, \widetilde{\Phi}_3\}$.  This is precisely the mirror model of the $N$HDM of Sec.~\ref{Sect:NHDM} after setting $N=3$.

Finally, we follow the same procedure to analyze the symmetry of $V_{\mathrm{3HDM}}^{(-)}$. Now, the operator $\alpha_A T^A$ must satisfy the relation from Eq.~(\ref{eq:Generator_relation}) for $S=S^-_a$, $a=1,\dots,6$ (see Eq.~\ref{eqn:smatrices}). The result is 
\begin{align}
\alpha_A T^A &=  \begin{pmatrix}
    \alpha_0 & 0 & 0 \\
    0 & \alpha_0 & 0 \\
    0 & 0 & \alpha_0
    \end{pmatrix}.
\end{align}
This means that the symmetry generators of $V_{\mathrm{3HDM}}^{(-)}$ are $\{\mathbb{1} \otimes \mathbb{1},  \mathbb{1} \otimes \dfrac{\sigma^a}{2}\}$, which indicates that the symmetry of the potential is $SU(2)\times U(1)=U(2)$. In this case, the constraints imposed by maximal entanglement do not enhance the symmetry of the potential.  

The existence of the solution $V_{\mathrm{3HDM}}^{(-)}$, as a stable potential that maximally entangles, challenges the proposed connection between maximal entanglement and enhanced symmetries.  To maintain this connection, we must look for other constraints to rule out such solutions.  One idea is to consider that the remaining symmetries have to be gauged.  As mentioned earlier, the two Higgs doublet model solution's $U(2)^2$ could be consistently gauged in a way where the new contributions from the gauge bosons still resulted in maximal entanglement for the Higgs scattering amplitudes.  For $V_{\mathrm{3HDM}}^{(+)}$, following the discussion in Sec.~\ref{Sect:NHDM}, if one gauges the $U(2)^3$ with identical couplings for the $SU(2)$ and $U(1)$ groups in the $\widetilde{\Phi}$ basis, maximal entanglement is maintained by the gauge boson exchange amplitudes.  However, for $V_{\mathrm{3HDM}}^{(-)}$, gauging the $U(2)$ no longer leads to maximal entanglement.  A simple scattering channel is to consider $H_1^+ H_2^0 \to H_k^+ H_l^0,$ with other channels having similar issues.  To simplify this, we will assume $U(2)$ is unbroken and the Higgs bosons are massless. The potential $V_{\mathrm{3HDM}}^{(-)}$ and the gauge interactions give a transition matrix element 
\begin{align}
|H_1^+ H_2^0\rangle \to \left(\frac{g'^2-g^2}{4t}(s-u)-2\lambda\right)|H_1^+  H_2^0\rangle +\left(\frac{g^2}{2u}(s-t)-2\lambda\right) |H_2^+ H_1^0\rangle + 2\lambda|H_3^+ H_3^0\rangle.
\end{align}
For this to  maximally entangle, one needs all three coefficients to have the same magnitude, so the first two coefficients must be equal to $\pm 2\lambda$.   For the second coefficient, since $\frac{s-t}{u}<0$, this requires $g=0$ so that $SU(2)$ is not actually gauged.  The first coefficient can equal $2\lambda$ if $i)\; g=g'$ or  if $ii)\; g' < g$ but only at a specific kinematic point.  Case $i)$ doesn't work for the second coefficient, while case $ii)$'s vanishing at a particular kinematic point is problematic in general.   In this example, since the potential contributions to the amplitude do not depend on $\cos\theta$, they give no insight into a general approach for how to handle this kinematic dependence, which will typically depend on both the center of mass energy and scattering angle.  At any rate, we see that adding the  additional constraint that maximal entanglement is maintained when the enhanced symmetry is gauged, could be a consistent prescription where max entanglement and stability of the theory can be reliably connected to enhanced gauge symmetries.  

\section{Conclusions}
\label{Sect:conclusions}
In this paper, we have taken an extended look at the potential connection between enhanced symmetries and maximizing flavor entanglement for Higgs doublet models with bounded potentials.  We have confirmed the analysis of \cite{Carena:2025wyh} for the 2HDM, finding consistency with their results, where the only maximal entangling model with bounded potential have mirror symmetries of $U(2)^2.$  Turning to $N$HDMs, we discovered a generalization of the mirror models, with enhanced symmetry of $U(2)^N.$

To determine if these are the only solutions, we looked in detail at the 3HDM, using a combination of numerical and analytical methods.  Here we found many maximally entangling solutions, but only two sets of solutions which are bounded from below.  One is indeed the realization of the $N$HDM mirror model, but surprisingly the other solution is stable but has only the minimal symmetry of $U(2).$  Thus, for the general connection between maximal entanglement and enhanced symmetries to be reliable, one must look for additional constraints.  For the 3HDM, we demonstrated that requiring maximal entanglement to be maintained after gauging the symmetries eliminated the solution without enhanced symmetries.  

To summarize, our work shows that the connection between enhanced symmetries and maximal entanglement needs refinement and could potentially require that entanglement maximization must be maintained even  after gauging the symmetries.  Given the ability of the mirror models to solve this issue, this leads to a natural conjecture that the maximally entangling solutions, which are stable and whose symmetries can be gauged, are models with mirror symmetry.  In the ``mirror" basis, these models have $N$ identical sectors which are decoupled from each other.  

Future work can help determine if this procedure works for more scenarios.  A natural extension is other minimal Higgs representations and seeing if the maximally entangling $N$ Higgs models indeed have enhanced symmetries which can be gauged.  If such consequences can be verified over many such examples, it provides stronger evidence for the general connection, which would motivate working towards a fundamental understanding of these, so far, empirical confirmations.  

\section*{Acknowledgments}
We thank I.~Low for useful discussions.  S.C.~and G.J.~are supported in part by the DOE under grant DE-SC0011640.

\section{References}\label{References}
\bibliographystyle{utphys}
\bibliography{references}

@article{Li:2026kha,
    author = "Li, Cihang and Ma, Teng and Shu, Jing and Zhu, Mingdi",
    title = "{Entanglement Maximization and Symmetry Selection in Composite Higgs Models}",
    eprint = "2605.17434",
    archivePrefix = "arXiv",
    primaryClass = "hep-ph",
    month = "5",
    year = "2026"
}

@article{Kanemura:2026won,
    author = "Kanemura, Shinya and Tanaka, Masanori",
    title = "{Entanglement entropy minimization and global symmetry violation in scatterings}",
    eprint = "2609.02691",
    archivePrefix = "arXiv",
    primaryClass = "hep-ph",
    reportNumber = "OU-HET 1323",
    month = "9",
    year = "2026"
}

@article{Liu:2025iwh,
    author = "Liu, Jia and Tanaka, Masanori and Wang, Xiao-Ping and Zhang, Jing-Jun and Zheng, Zifan",
    title = "{Parameter inference from final-state entanglement in Higgs boson decays}",
    eprint = "2511.17321",
    archivePrefix = "arXiv",
    primaryClass = "hep-ph",
    reportNumber = "CPTNP-2025-042",
    doi = "10.1103/jynb-w9kq",
    journal = "Phys. Rev. D",
    volume = "114",
    number = "1",
    pages = "015036",
    year = "2026"
}

@article{Sone:2026jmo,
    author = "Sone, Katsuyoshi and Hu, Tao-Ran and Guo, Feng-Kun and Hyodo, Tetsuo and Low, Ian",
    title = "{Entanglement suppression for {\ensuremath{\Omega}}{\ensuremath{\Omega}} scattering}",
    eprint = "2602.09630",
    archivePrefix = "arXiv",
    primaryClass = "hep-ph",
    reportNumber = "FERMILAB-CONF-26-0174-V",
    doi = "10.1142/s0217732326420071",
    journal = "Mod. Phys. Lett. A",
    volume = "41",
    number = "30n31",
    pages = "2642007",
    year = "2026"
}

@article{Kowalska:2025qmf,
    author = "Kowalska, Kamila and Sessolo, Enrico Maria",
    title = "{Qubit entanglement from forward scattering}",
    eprint = "2510.04200",
    archivePrefix = "arXiv",
    primaryClass = "hep-ph",
    doi = "10.1007/JHEP04(2026)014",
    journal = "JHEP",
    volume = "04",
    pages = "014",
    year = "2026"
}

@article{Carena:2025wyh,
    author = "Carena, Marcela and Coloretti, Guglielmo and Liu, Wanqiang and Littmann, Mira and Low, Ian and Wagner, Carlos E. M.",
    title = "{Entanglement maximization and mirror symmetry in two-Higgs-doublet models}",
    eprint = "2505.00873",
    archivePrefix = "arXiv",
    primaryClass = "hep-ph",
    reportNumber = "FERMILAB-PUB-25-0308-V",
    doi = "10.1007/JHEP08(2025)016",
    journal = "JHEP",
    volume = "08",
    pages = "016",
    year = "2025"
}

@article{McGinnis:2025brt,
    author = "McGinnis, Navin",
    title = "{Symmetry, entanglement, and the $S$-matrix}",
    eprint = "2504.21079",
    archivePrefix = "arXiv",
    primaryClass = "hep-th",
    month = "4",
    year = "2025"
}

@article{Thaler:2024anb,
    author = "Thaler, Jesse and Trifinopoulos, Sokratis",
    title = "{Flavor patterns of fundamental particles from quantum entanglement?}",
    eprint = "2410.23343",
    archivePrefix = "arXiv",
    primaryClass = "hep-ph",
    reportNumber = "MIT-CTP 5788",
    doi = "10.1103/PhysRevD.111.056021",
    journal = "Phys. Rev. D",
    volume = "111",
    number = "5",
    pages = "056021",
    year = "2025"
}

@article{Cervera-Lierta:2017tdt,
    author = "Cervera-Lierta, Alba and Latorre, Jos\'e I. and Rojo, Juan and Rottoli, Luca",
    title = "{Maximal Entanglement in High Energy Physics}",
    eprint = "1703.02989",
    archivePrefix = "arXiv",
    primaryClass = "hep-th",
    doi = "10.21468/SciPostPhys.3.5.036",
    journal = "SciPost Phys.",
    volume = "3",
    number = "5",
    pages = "036",
    year = "2017"
}

@article{Chang:2024wrx,
    author = "Chang, Spencer and Jacobo, Gabriel",
    title = "{Consequences of minimal entanglement in bosonic field theories}",
    eprint = "2409.13030",
    archivePrefix = "arXiv",
    primaryClass = "hep-ph",
    doi = "10.1103/PhysRevD.110.096020",
    journal = "Phys. Rev. D",
    volume = "110",
    number = "9",
    pages = "096020",
    year = "2024"
}

@article{Low:2021ufv,
    author = "Low, Ian and Mehen, Thomas",
    title = "{Symmetry from entanglement suppression}",
    eprint = "2104.10835",
    archivePrefix = "arXiv",
    primaryClass = "hep-th",
    doi = "10.1103/PhysRevD.104.074014",
    journal = "Phys. Rev. D",
    volume = "104",
    number = "7",
    pages = "074014",
    year = "2021"
}

@article{Liu:2022grf,
    author = "Liu, Qiaofeng and Low, Ian and Mehen, Thomas",
    title = "{Minimal entanglement and emergent symmetries in low-energy QCD}",
    eprint = "2210.12085",
    archivePrefix = "arXiv",
    primaryClass = "quant-ph",
    doi = "10.1103/PhysRevC.107.025204",
    journal = "Phys. Rev. C",
    volume = "107",
    number = "2",
    pages = "025204",
    year = "2023"
}

@article{Beane:2018oxh,
    author = "Beane, Silas R. and Kaplan, David B. and Klco, Natalie and Savage, Martin J.",
    title = "{Entanglement Suppression and Emergent Symmetries of Strong Interactions}",
    eprint = "1812.03138",
    archivePrefix = "arXiv",
    primaryClass = "nucl-th",
    reportNumber = "INT-PUB-18-056, NT@UW-18-19",
    doi = "10.1103/PhysRevLett.122.102001",
    journal = "Phys. Rev. Lett.",
    volume = "122",
    number = "10",
    pages = "102001",
    year = "2019"
}

@article{Hu:2024hex,
    author = "Hu, Tao-Ran and Chen, Su and Guo, Feng-Kun",
    title = "{Entanglement suppression and low-energy scattering of heavy mesons}",
    eprint = "2404.05958",
    archivePrefix = "arXiv",
    primaryClass = "hep-ph",
    doi = "10.1103/PhysRevD.110.014001",
    journal = "Phys. Rev. D",
    volume = "110",
    number = "1",
    pages = "014001",
    year = "2024"
}

@article{Carena:2023vjc,
    author = "Carena, Marcela and Low, Ian and Wagner, Carlos E. M. and Xiao, Ming-Lei",
    title = "{Entanglement suppression, enhanced symmetry, and a standard-model-like Higgs boson}",
    eprint = "2307.08112",
    archivePrefix = "arXiv",
    primaryClass = "hep-ph",
    reportNumber = "FERMILAB-PUB-23-269-T",
    doi = "10.1103/PhysRevD.109.L051901",
    journal = "Phys. Rev. D",
    volume = "109",
    number = "5",
    pages = "L051901",
    year = "2024"
}

@article{Kowalska:2024kbs,
    author = "Kowalska, Kamila and Sessolo, Enrico Maria",
    title = "{Entanglement in flavored scalar scattering}",
    eprint = "2404.13743",
    archivePrefix = "arXiv",
    primaryClass = "hep-ph",
    doi = "10.1007/JHEP07(2024)156",
    journal = "JHEP",
    volume = "07",
    pages = "156",
    year = "2024"
}

\end{document}